\documentclass{iau}
\usepackage{graphicx}

\title[IAUS291.~~Magnetar versus Quarctar] 
{AXPs \& SGRs: Magnetar or Quarctar?}

\author[G.J. Qiao et al.]   
{Guojun Qiao$^1$, Xionwei Liu$^1$, Renxin Xu$^1$, Yuanjie Du$^2$,
Jinlin Han$^3$, Hao Tong$^4$
 \and Hongguang Wang$^5$}

\affiliation{$^1$Department of Astronomy, Peking University,
 Beijing 100871, China, \\ email: {\tt gjn@pku.edu.cn, xiongwliu@163.com, r.x.xu@pku.edu.cn} \\[\affilskip]
$^2$Center for Space Science and Applied Research, CAS, Beijing,
100190, China \\email: {\tt duyj@nssc.ac.cn}\\[\affilskip]
 $^3$National Astronomical Observatories, CAS, Beijing 100012, China \\email:
{\tt hjl@nao.cas.cn}\\[\affilskip]
 $^4$Xinjiang Astronomical Observatory, CAS, Urumqi, Xinjiang 830011, China \\email: {\tt
tonghao@xao.ac.cn}\\[\affilskip]
$^5$Center for Astrophysics, Guangzhou University,
Guangzhou 510400, China \\email: {\tt cosmic008@yahoo.com.cn}}

\pubyear{2012}
\volume{291}  
\jname{\mbox{Neutron Stars and Pulsars: Challenges and Opportunities after 80 years}}
\editors{J. van Leeuwen, ed.}
\begin{document}

\maketitle

\begin{abstract}
The concept of a ``magnetar" was proposed mainly because of two factors.
First, the X-ray luminosity of Anomalous X-ray Pulsars (AXPs) and
Soft Gamma-Ray Repeaters (SGRs) is larger than the rotational
energy loss rate ($Lx > \dot E_{rot}$), and second, the magnetic
field strength calculated from ``normal method" is super strong. It is proposed that the radiation energy of magnetar comes from its magnetic fields. Here it is argued that
the magnetic field strength calculated through the normal method is
incorrect at the situation $Lx > \dot E_{rot}$, because the wind braking
is not taken into account. Besides, the ``anti-magnetar" and some other X-ray and
radio observations are difficult to understand with a magnetar model.

Instead of the magnetar, we propose a ``quarctar", which is a
crusted quark star in an accretion disk, to explain the observations. In this model, the
persistent X-ray emission, burst luminosity, spectrum of AXPs and
SGRs can be understood naturally. The radio-emitting AXPs, which are challenging the magnetar, can also be explained by the quarctar model.

\end{abstract}

\firstsection 
\section{Introduction}
Anomalous X-ray Pulsars (AXPs) and Soft Gamma-Ray Repeaters (SGRs)
have been generally recognized as neutron stars with super strong
magnetic fields, namely magnetars. The primary properties of
magnetars are: \\
1) The strength of magnetic field calculated from
normal method is super-strong; \\
2) The X-ray luminosities are
larger than the rotational energy loss rate, i.e. $L_{X}>\dot
{E}_{\rm rot}$;  \\
3) The radiation energy comes from the energy
of the magnetic field.

Here it is argued that: 1) The magnetic field can not be calculated
from normal method when $L_{X}>\dot {E}_{\rm rot}$; 2)
$L_{X}>\dot {E}_{\rm rot}$ does not mean that the radiation energy is
coming from the energy of the magnetic field; 3) Pulsed radio
emission observed from some AXPs implies that it should be
found another way to understand the observed phenomenons.

Following the arguments we present a model, quarctar, whose main properties are described below
briefly.

\section{Do magnetars really exist?}
\subsection{Is the magnetic field strength calculated from normal method correct?}
When considering the following points, one finds that the method to calculate the
magnetic field widely used in the literature is not suitable for
AXP/SGR (Qiao et al. 2010). \\
1) The energy loss is carried out not only by magnetic dipole
radiation but also by wind; \\
2) Particle acceleration and
radiation depend on the electric potential in the polar cap
regions; \\
3) If the energy carry out by the high energy particles is
larger than the dipole radiation, then the value of magnetic field
calculated from the classical method, i.e., the rotational energy loss
equals to the dipole magnetic radiation energy loss, $\dot
{E}_{rot}= \dot {E}_{dipole}$, is incorrect. The wind braking should
be taking into account (Tong et al. 2012). Probably, this is just
the case of AXPs/SGRs.

\subsection{Anti-magnetar observations: challenge to magnetars}

There are observed anti-magnetar phenomena.
For example, in PSR J1852+0040, $P=105$ ms, $\dot{P} = (8.68\pm
0.09)\times 10^{-18}~\rm \,s\,s^{-1}$, the surface magnetic field is
$B_{s} = 3.1\times10^{10}$\,G, which is the weakest magnetic field
ever measured for a young neutron star. Its X-ray luminosity is $L_{X}
= 5.3\times 10^{33}(d/7.1$ kpc$)^2$\,erg\,s$^{-1}$, while the rotational
energy loss is $\dot{E}_{\rm rot} = 3.0 \times
10^{32}$\,erg\,s$^{-1}$; thus $L_{X}/\dot{E}_{\rm rot}\simeq 17.7$
(Halpern \& Gotthelf 2010). This means that: $L_{x}>\dot{E}_{rot}$, which
does not mean super-strong magnetic field at all! Beside that, some observations show that weak field ``anti-magnetar'' neutron star
is still not ruled out for SN1987A (Manchester 2007; Gotthelf \&
Halpern 2008).

\subsection{Radio observations: the difference between radio pulsars and
magnetars does not originate from the difference of magnetic
fields}

Previous observations mainly show that: (1) no radio emission from
 magnetars are observed; (2) the magnetic fields of magnetars
 are stronger than those of normal radio pulsars.  Then it is
 generally believed that these differences are caused by the
 difference of the magnetic fields.

Recent observations show that all these two differences are
confusing: some radio pulsars have stronger magnetic field than that
of some AXPs (such as Hobbs et al. 2004; Kaspi \& McLaughlin
2005); radio emission from some AXPs are observed clearly after
X-ray flare (Halpern et al. 2005; Camilo et al. 2006; Lazaridis et
al. 2008). These mean that the differences between radio pulsars and
magnetars do not originate from the difference of magnetic
fields strength.

\section{AXPs \& SGRs are quarctars?}

Instead of the magnetars,  we
suggested a quarctar model: a quark star with a crust in an
accretion disk to account for the X-ray and radio emission properties.

\subsection{Quark star with crust: X-ray emission}

It is assumed that crust of quark stars would be
formed after a supernova explosion (e.g. Alcock et al. 1986). This
kind of quark stars can not be observed as radio pulsars, but bare
quark stars can be observed as radio pulsars (Xu et al. 2001).  Therefore
quark star with crust  may be observed in X-ray bands. The pulse
profiles in X-ray light curves should be wider and  consistent with
observations.

\subsection{Quark star with two polar holes: radio emission}

In the super flare,  owning to the out-flow of high energy particles from the polar
cap regions, two holes will form in the crust after some time. In this case, the polar cap regions become ``bare'', so that it can generate radio emission from the bared regions. Since the quark star lies in an accretion disk, later, the polar
holes will be filled by the accreted matter, hence it becomes radio quite. This scenario is consistent with
observations.

\subsection{Energy source of the radiations}

{\it Phase transition energy } comes from the phase transition of
normal matter to strange matter that takes place near the polar cap
regions of the strange stars with a crust. For a magnetized star,
the accreting material falls to the surface  through the
magnetic tube along the open field lines. The ions can be formed at
the polar cap regions. The ions  can be supported by the electromagnetic
force.  However, this balance can be destroyed easily,  so that the normal matter can be
transferred to the quark matter. This is a way to support  the energy loss from
the polar cap regions.

Xu et al. (2006) suggested
that the energy of superflares of SGR can  be supported by giant quakes in
solid quark stars,  which is a rich energy source for SGRs.

\begin{figure}[t]
\begin{center}
 \includegraphics[width=3.4in]{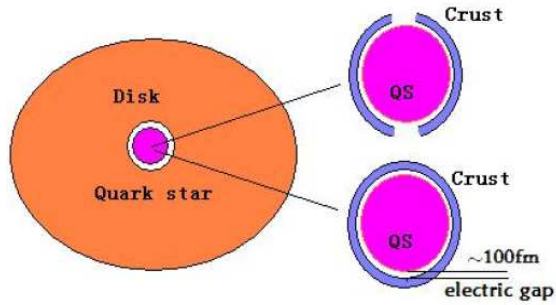}
 \caption{A quark star with a crust in an accretion disk, namely quarctar.
  Around the quark star there is a crust, in normal case, we can just observe
  radiation in X-ray bands; in the super flare, high energy particles flowing out
  from the pole cap regions, so there are two polar cap holes formed in the crust,
  in this case one can observe radio emission from polar cap regions. }
   \label{fig1}
\end{center}
\end{figure}

\acknowledgments
This work is supported by the National Basic Research Program of
China (2009CB824800 and 2012CB821800), the NSFC (10935001,
10973002,10833003,11103021,10821061, 10573002, 10778611, 10773016,
11073030) and the Key Grant Project of Chinese Ministry of Education
(305001).

\end{document}